\documentclass[12pt]{article}
\usepackage{subfigure}
\usepackage{amssymb,amsmath}
\usepackage{graphicx}
\usepackage{color}
\usepackage{epsfig}
\usepackage[colorlinks=true
,urlcolor=blue
,citecolor=blue
,linkcolor=blue
,pagecolor=blue
,linktocpage=true
,pdfproducer=medialab
]{hyperref}
\usepackage[a4paper,width=15.2cm]{geometry}
\makeatletter \renewcommand{\@dotsep}{10000} \makeatother
\usepackage{appendix}

\newcommand{\beq}{\begin{equation}}
\newcommand{\eeq}{\end{equation}}
\newcommand{\bea}{\begin{eqnarray}}
\newcommand{\eea}{\end{eqnarray}}

\begin{document}

\begin{center}

 {\Large\bf  Numerical Methods and Causality in Physics

 } \vspace{1cm}

{\large   Muhammad Adeel Ajaib\footnote{ E-mail: adeel@udel.edu}}

{\baselineskip 20pt \it
University of Delaware, Newark, DE 19716, USA  } \vspace{.5cm}

\vspace{1.5cm}
\end{center}

\begin{abstract}

We discuss physical implications of the explicit method in numerical analysis. Numerical methods have there own condition for causality, known as the Courant-Friedrichs-Lewy condition. It is proposed that numerical causality merges with physical causality as the grid interval size approaches zero. We discuss the implications of this proposition on the numerical analysis of the wave equation. We also show that, insisting on physical causality, the numerical analysis of Schrodinger's equation implies that the minimum space interval should satisfy $\Delta x \ge a_0 \lambda_c$, where $\lambda_c$ is the reduced Compton wavelength and $a_0$ is a constant of the order unity.

\end{abstract}

\newpage

\section{Introduction}\label{intro}

Partial differential equations (PDEs) are ubiquitous in all fields of science and are  used to model physical systems. In numerous cases, obtaining the exact solutions of these equations is not possible and one has to resort to numerical approximations in order to solve these systems. Numerical methods therefore have a wide range of applications in numerous scientific fields. In physics, for instance, these methods are used to obtain approximate solutions of important differential equations such as the diffusion and wave equation.

In this article we attempt to demonstrate how a formalism used to solve PDEs in applied mathematics can be employed to further our understanding of known fundamental ideas in physics. 
The numerical solutions of widely used equations in physics, such as the wave and heat equation, are known and there are different methods to solve them. We shall use the explicit finite difference method to obtain solutions of the wave and Schrodinger equations. We will use the Von Neumann stability analysis to discuss the limits on the grid intervals. This analysis typically leads to limits on the space and time intervals of the numerical grid in order to ensure stability. We will discuss these limits while emphasizing on physical causality. 

The notion of causality is one of the foundational ideas in modern physics. In the context of Einstein's special theory of relativity it implies a Universal limit on the speed with which cause and effect are related. 
This universal limit being the speed of light. It is the limiting speed with which information can travel in the Universe. 
 In other words, causality implies that the effect of an event in space-time should belong to its future light cone. The stability requirement of the explicit method in numerical analysis typically leads to a condition of causality, known as the Courant-Friedrichs-Lewy (CFL)
 condition. We propose in this article that in the continuum limit numerical causality unifies with physical causality.  We will show that insisting on physical causality in the explicit method of the Schrodinger equation can lead to limits on the minimum grid interval sizes which are essentially consistent with those obtained in quantum mechanics.

The paper is organized as follows: In section \ref{intro-num-analysis} we briefly introduce the finite difference methods used in numerical analysis to solve partial differential equations. Section \ref{sec:wave-eq} includes a discussion on the wave equation and the stability condition which implies an upper limit on the speed of waves. In section \ref{sec:schrodinger} we show that  the convergence and causality in the explicit method of the Schrodinger equation imply the minimum grid interval size to be of the order of Compton wavelength.  Section \ref{numerical-discussion} includes a general discussion on the properties of the explicit method that accord it a causal structure. We conclude in section \ref{conclude}.

\section{Numerical analysis and discretization} \label{intro-num-analysis}
Finite difference methods (FDMs) are frequently used in numerical analysis to solve partial differential equations. In these methods a PDE is solved on a grid of discrete points with, for instance, $x_n=n \Delta x$ and $t_j=j \Delta t$, where $n$ and $j$ are integers \cite{Gilbert:2006, Gilbert:2007, Morton-Mayers}. Continuous derivatives are then replaced by finite difference approximations. The solution of complex differential equations can be obtained by dividing space-time into a discrete set of points in this manner. The finite difference equation, for example, can be a forward, backward or centered difference. Each choice needs to be tested in various ways such as its stability, accuracy and speed. 

A numerical method is stable if the errors in the method do not increase as the solution moves forward in time. There can be several sources of errors in a numerical method. These include, for example,  the truncation 
and round-off errors. The round-off error results from finite precision representation of real numbers on a computer. Truncation errors arise from representing continuous derivatives with finite differences which involves truncating terms from the Taylor series. The leading terms in the truncation error determines the accuracy of a FDM.

What makes a particular numerical method feasible is its consistency, i.e., whether the finite difference equation (FDE) approaches the PDE as the interval size vanishes. In other words, a numerical scheme is consistent if the truncation error goes to zero as $\Delta x, \Delta t \rightarrow 0$. Similarly, convergence is also important and requires the solution of the FDE to converge to the exact solution in the limit $\Delta x, \Delta t \rightarrow 0$. According to the Lax equivalence theorem, if a finite difference method is consistent (FDE approaches the PDE as grid interval size approaches zero) and stable (errors do not grow) than convergence (discrete solution converges to the actual solution as grid interval size approaches zero) is guaranteed. In short, consistency and stability imply convergence.

The most convenient way to test the stability of a finite difference method is by employing the Von Neumann stability analysis. In this method we test how a Fourier mode behaves on the grid. Compared to other methods, such as the matrix method, this analysis is a convenient way to test the stability of a method. One short coming of this method is that it ignores the boundary conditions. This is why the limits obtained are necessary but not sufficient to guarantee stability.
 But this short coming, in a way, is useful in our case since the limits we get would be independent of the boundary conditions of the problem. In FDMs, a first order derivative in space and time can be approximated by a forward time and space difference as follows
\begin{eqnarray}
\frac{\partial u}{\partial x} \simeq \frac{ u_{j+1}- u_j}{\Delta x}, \\
\frac{\partial u}{\partial t} \simeq \frac{ u^{n+1}- u^{n}}{\Delta t},
\end{eqnarray}
where $i=0,\ 1, 2 \ldots M$ and $n=0,\ 1, 2 \ldots N$.   
Amongst several methods used, two important ones are the explicit and implicit methods. Each of these has its own benefits in terms of stability, speed and accuracy of the solution. The explicit method, which is a forward difference method, is fast but less stable since it usually requires limits on the space and time step sizes. In the explicit method the state of the system at a time $t_{n+1}$ is evaluated from the state at time $t_n$ or a prior time ($t_{n-1},..$), namely
\begin{eqnarray}
u_{n+1}=f(u_{n},t_n).
\label{eq:explicit1}
\end{eqnarray}

The implicit method, which is a backward difference method, is slow and more numerically stable. In this method the state of the system at a later time is not an explicit function of the state at earlier times.  For this case, therefore, the state of the system at a later time would typically be given by
\begin{eqnarray}
u_{n+1}=f(u_{n},u_{n+1},t_{n+1}).
\label{eq:implicit1}
\end{eqnarray}
 The factor that makes this method more stable is that the time step is usually not constrained by a stability condition. A larger time step $\Delta t$ can therefore be used to solve the system.  However the above equation is usually a non-linear system of equations and needs additional time to solve. Therefore this method is slower than the explicit method. We will focus on the explicit method in the following sections.

\begin{figure}[t!]
\begin{center}
\includegraphics[scale=1.2]{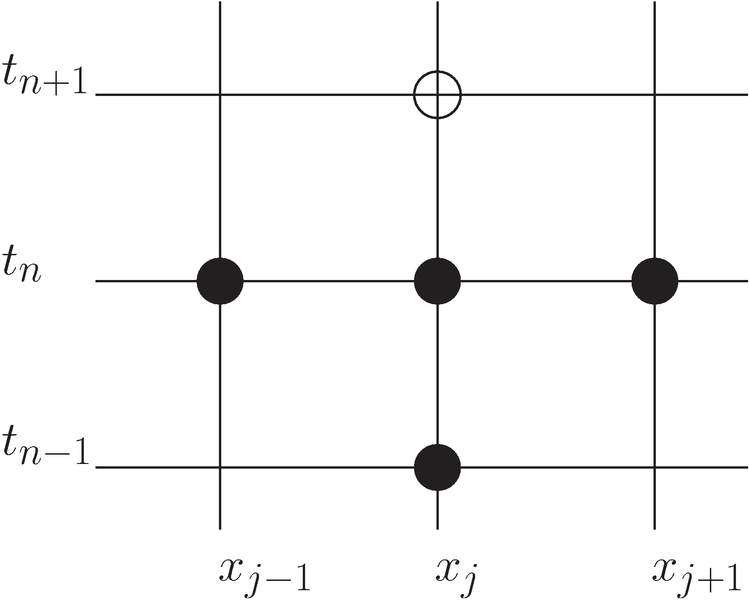}
\end{center}
\caption{Stencil for the FDM of the 1-D wave equation. It represents the expression given in equation (\ref{eq:wave-eq3}).}
\label{fig:wave-eq}
\end{figure}



\section{Wave equation: Speed Limit}\label{sec:wave-eq}

In this section, we show that employing the explicit finite difference method in solving the wave equation can lead to a limit on the propagation speed of waves. The  wave equation is a hyperbolic PDE and describes the propagation of waves. Hyperbolic PDEs describe systems with a finite speed of propagation.  Consider the one dimensional wave equation:
\begin{eqnarray}
\frac{\partial^2 u}{\partial t^2} = v^2 \frac{\partial^2 u}{\partial x^2} .
\label{eq:wave-eq1}
\end{eqnarray}
We can solve the wave equation using the explicit method by replacing the derivatives by centered differences in space and time (leapfrog method) as follows
\begin{eqnarray}
\frac{\partial^2 u}{\partial x^2} = \frac{u_{j+1}^{n}-2 u_{j}^{n}+u_{j-1}^n}{\Delta x^2}+O(\Delta x^2) , \\
\frac{\partial^2 u}{\partial t^2} = \frac{u_{j}^{n+1}-2 u_{j}^{n}+u_{j}^{n-1}}{\Delta t^2}+O(\Delta t^2) ,
\label{eq:wave-eq2}
\end{eqnarray}
where $u_j^n \equiv u(x_j, t_n)$. The truncation error in this case is $O(\Delta x^2)$ and $O(\Delta t^2)$ and arises from truncating terms from the Taylor expansion. The leading terms in the truncation error of this method are given by
\begin{eqnarray}
T^n_j = \frac{1}{12} v^2 (\Delta x)^2 \ u_{xxxx}-\frac{1}{12} (\Delta t)^2 \ u_{tttt}.
\label{eq:wave-eq2b}
\end{eqnarray}
Dropping the truncation error we get the following approximation
\begin{eqnarray}
u_{j}^{n+1}=r^2(u_{j+1}^{n}-u_{j-1}^n)+2(1-r^2)u_{j}^{n}-u_{j}^{n-1},
\label{eq:wave-eq3}
\end{eqnarray}
where $r=v \Delta t / \Delta x$. 
To study the stability of this method we use the Von Neumann or Fourier analysis. We look for solutions of the form 
\begin{eqnarray}
u^n_j=e^{i k (j \Delta x) } G^n( \Delta t, \Delta x ,k), 
\label{eq:schrodinger-fd3}
\end{eqnarray}
 and get the following equation for the growth factor
\begin{eqnarray}
G^2-2[1-2r^2 \sin ^2(k \Delta x)]G+1=0 .
\label{eq:wave-eq4}
\end{eqnarray}
The requirement that $|G| \le 1$ leads to the following CFL condition for stability
\begin{eqnarray}
v \leq  \frac{\Delta x}{\Delta t} . 
\label{eq:stab-cond1}
\end{eqnarray}
 In 3 dimensions, the CFL condition for stability is given by \cite{Gilbert:2007} 
\begin{eqnarray}
v \le \frac{1}{\Delta t} \left[ \frac{1}{\Delta x^2}+\frac{1}{\Delta y^2}+\frac{1}{\Delta z^2}\right]^{-1/2}  \equiv \Delta v_{grid},
\label{eq:stab-cond2a}
\end{eqnarray}
which  for a cubic grid ($\Delta x=\Delta y=\Delta z$) will be
\begin{eqnarray}
v \leq \frac{1}{\sqrt{3}} \frac{\Delta x}{\Delta t} \equiv \Delta v_{grid}. 
\label{eq:stab-cond2}
\end{eqnarray}
Here $ \Delta v_{grid}$ is the average grid velocity and can be viewed as the speed with which information moves across the grid \cite{Holmes:2006}. We can view equation (\ref{eq:stab-cond1}) as a limit on the velocity of the wave. It requires the domain of dependence of the wave equation to be within the numerical domain of dependence. The question of making $\Delta x$ and $\Delta t$ smaller is an important one in numerical analysis. As $\Delta t$ is made smaller the time for
 the computation increases rapidly. So, practicality also dictates $\Delta t$ not to be too small.
 
 As described in section \ref{intro-num-analysis}, in order for the explicit method to be feasible it should also be convergent. The truncation error for this method given in equation (\ref{eq:wave-eq2b}) is $O(\Delta x^2, \Delta t^2)$ and vanishes in the limit $\Delta x, \Delta t \rightarrow 0$.  However, as the grid interval size approaches zero we need to specify the relationship between $\Delta x$ and $\Delta t$. We approach $\Delta x, \Delta t \rightarrow 0$ such that the ratio $\Delta x/ \Delta t$ is kept constant. This is referred to as the \textit{refinement path} \cite{Morton-Mayers}.  
%

\begin{figure}[t!]
\begin{center}
\includegraphics[scale=.55]{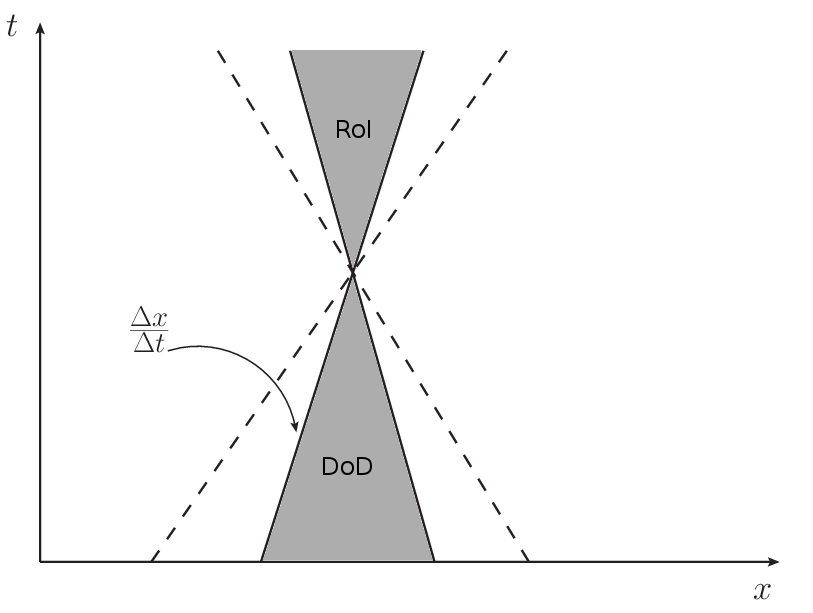}
\end{center}
\caption{Figure shows the domain of dependence and range of influence for the 1-D wave equation. The gray region bounded by the solid line represents the continuous PDE whereas the dashed line shows the boundary of the numerical solution. The CFL condition in equation (\ref{eq:stab-cond1}) requires the DoD of the PDE to be contained within the DoD of the numerical solution. As the grid interval size approaches zero the slope of the dashed boundary representing the numerical PDE should approach the limiting speed with which information can be transferred across the grid. So in the continuum limit, $v \le c$, if we insist that numerical causality merges with physical causality.} 
\label{fig:dod}
\end{figure}


Furthermore, the speed with which numerical information for the explicit method flows on the grid is different from the physical speed of information transfer.   
Herein we will assume that the numerical speed of information becomes equal to the physical speed of information as the grid approaches the continuum limit. In other words, numerical causality merges with physical causality in the continuum limit.
Therefore, taking the limit $\Delta x, \Delta t \rightarrow 0$ while keeping the instantaneous velocity $v_{grid}$ constant we get
\begin{eqnarray}
v &\leq &  v_{grid} = c, \nonumber \\
\Rightarrow v & \leq &   c,
\label{eq:stab-cond3}
\end{eqnarray}  
where the constant $c$ is the speed of light. It is the limiting speed with which information travels in the continuum limit. Therefore, by assuming the numerical speed to be the same as the physical speed limit as $\Delta x, \Delta t \rightarrow 0$, we get a limit on the speed of waves which obeys causality. This limit also ensures the stability of the explicit method.

Note that the numerical solution of Maxwell's equation is obtained using the Finite Domain Time Difference (FDTD) method and results in a conditional stability limit similar to equation (\ref{eq:stab-cond2a}) (see, for example, chapter 4 of ref. \cite{Taflove}). For the case of Maxwell's equation the condition (\ref{eq:stab-cond2a}) becomes an equality in the continuum limit.

\textit{Causality:} The domain of dependence (DoD) and range of influence (RoI) of a hyperbolic PDE is shown in Fig. \ref{fig:dod}. The DoD of a point in the solution domain is the set of points on which the solution at a particular point depends. Similarly, the RoI is the solution domain which is affected by the solution at a particular point. The CFL condition in equation (\ref{eq:stab-cond1}) also ensures that the PDE domain of dependence stays within the numerical domain of dependence as shown if Fig. \ref{fig:dod}. The gray region is the DoD of the PDE which is contained in the DoD of the numerical solution.
The numerical value of the solution at a node $(x_i,t_n)$ depends on the values at the nodes lying within the numerical domain of dependence.   
The wave equation therefore obeys causality and, particularly, in the continuum limit, equation  (\ref{eq:stab-cond3}) guarantees that physical causality is always satisfied.


\section{Schrodinger equation: Minimum Space and time intervals}\label{sec:schrodinger}

In this section, we show that insisting on physical causality and convergence in the explicit finite difference method to solve the Schrodinger equation yields a limit on the minimum grid  interval size. 
The Schrodinger equation can be obtained from the non-relativistic limit of the relativistic expression for energy, $E^2=\vec{p}^{\, 2} + m^2$, by replacing relevant variables by operators as, $E-m \simeq  \hat{ H} =i\hbar \partial /\partial t$ and $\hat{p} = \hbar/i \ \partial /\partial x$. The time dependent Schrodinger equation for a free particle in one dimension is given by
\begin{eqnarray}
\frac{\partial \psi}{\partial t} = \frac{i \hbar}{2 m} \frac{\partial^2 \psi}{\partial x^2}.
\label{eq:schrodinger}
\end{eqnarray}
The Schrodinger equation is similar to the diffusion equation but with an imaginary diffusion coefficient. This changes the properties of the solutions completely. Schrodinger's equation allows plane wave solution whereas the solutions of the diffusion equation are diffusive. We first employ the finite difference form of the Schrodinger equation (\ref{eq:schrodinger}) with forward difference in time and centered difference in space (FTCS) and show that it is unstable. Therefore, equation (\ref{eq:schrodinger}) can be approximated as
\begin{eqnarray}
\frac{ \psi_j^{n+1}-\psi_j^{n}}{\Delta t} = \frac{i \hbar}{2 m} \frac{\psi_{j+1}^{n}-2 \psi_{j}^{n}+\psi_{j-1}^n}{(\Delta x)^2}+ O(\Delta t, \Delta x^2),
\label{eq:schrodinger-fd}
\end{eqnarray}
which yields
\begin{eqnarray}
\psi_j^{n+1} =  \psi^n_j + i   \alpha  \beta(\psi_{j+1}^{n}-2 \psi_{j}^{n}+\psi_{j-1}^n),
\label{eq:schrodinger-fd2}
\end{eqnarray}
where, $\beta=\hbar/m$ and $\alpha = \Delta t/2(\Delta x)^2$. The leading terms in the truncation error of this method are given by
\begin{eqnarray}
T^n_j =  \frac{1}{2} \Delta t \ \psi_{tt}- \frac{i \hbar}{2 m} \frac{1}{12} (\Delta x)^2 \ \psi_{xxxx}.
\label{eq:schrodinger-fd2b}
\end{eqnarray}
Consistency requires this error to vanish as the grid approaches the continuum limit. 
To find the growth factor $G=G( \Delta t,\Delta x ,k)$ we use the Von Neumann analysis by
 substituting the following expression in equation (\ref{eq:schrodinger-fd2})
\begin{eqnarray}
\psi^n_j=e^{i k (j \Delta x) } G^n( \Delta t, \Delta x ,k), 
\label{eq:schrodinger-fd3}
\end{eqnarray}
and attain the following equation for the one step growth factor $G$  
\begin{eqnarray}
G=1+2 i  \alpha \beta [\cos (k \Delta x)-1]. 
\label{eq:schrodinger-fd4}
\end{eqnarray}
The condition for stability is that $|G| \le 1$  should be true for all $k \Delta x$. 
The magnitude of the growth factor in equation (\ref{eq:schrodinger-fd4}) is 
\begin{eqnarray}
|G|^2=1+[2  \beta \alpha (\cos (k \Delta x)-1)]^2 \ge 1. 
\label{eq:schrodinger-fd4b}
\end{eqnarray}
Therefore, the growth factor is greater than 1 for all $k \Delta x \neq 0$ and this method is unstable.
The FTCS method along with some other explicit schemes to solve the Schrodinger's equation are known to be unstable \cite{Chan-lee}. For this reason the implicit methods are typically employed to solve it. The fully implicit method for the Schrodinger equation involves the centered difference in space, i.e. the right hand side of equation (\ref{eq:schrodinger-fd}), at time $t_{n+1}$. In the Crank Nicholson method the average of the explicit and implicit schemes is taken and this method is known to be unconditionally stable for the Schrodinger's equation \cite{Goldberg}. Although stable, this method is computationally  expensive as it involves inversion of large matrices. Moreover, it has infinite speed of propagation of numerical information and is therefore not suitable if we insist on physical causality in the continuum limit. As mentioned earlier, this is the reason we will focus on the explicit method. 

One of the reasons for the instability of the FTCS method is that it is not centered in time \cite{Leforestier}. To overcome this an explicit finite difference scheme was proposed in \cite{Askar} which is centered in time, unitary and conditionally stable. The following approximation is taken in this method
\begin{eqnarray}
\frac{ \psi_j^{n+1}-\psi_j^{n-1}}{2\Delta t} = \frac{i \hbar}{2 m} \frac{\psi_{j+1}^{n}-2 \psi_{j}^{n}+\psi_{j-1}^n}{(\Delta x)^2} .
\label{eq:schrodinger-fd5}
\end{eqnarray}
The stencil for this equation is similar to the one for the wave equation shown in Fig. \ref{fig:wave-eq}. The truncation error for this case is $O(\Delta t^2, \Delta x^2)$. The leading terms in the truncation error of this method are given by
\begin{eqnarray}
T^n_j =  \frac{1}{6} (\Delta t)^2 \ \psi_{ttt}- \frac{i \hbar}{2 m} \frac{1}{12} (\Delta x)^2 \ \psi_{xxxx}.
\label{eq:schrodinger-fd2b}
\end{eqnarray}
The Von Neumann stability analysis leads to the following equation for the growth factor 
\begin{eqnarray}
G^2 +4 i  \alpha \beta [1- \cos (k \Delta x)] G-1=0. 
\label{eq:schrodinger-fd5b}
\end{eqnarray}
The growth factors are therefore
\begin{eqnarray}
G_{1,2}=-2 i \alpha \beta [1- cos(k \Delta x)] \pm \sqrt{1-4 \alpha^2 \beta^2 [1-cos(k \Delta x)]^2}.
\label{eq:schrodinger-fd5c}
\end{eqnarray}
The requirement $|G| \le 1$ leads to the following stability condition \cite{Askar, Harmuth}
\begin{eqnarray}
\alpha \beta \le \frac{1}{4},
\end{eqnarray}
which implies
\begin{eqnarray}
\frac{\Delta x^2}{ \Delta t} \ge \frac{2\hbar}{m}.
\label{eq:schrodinger-fd5}
\end{eqnarray}
The stability condition in 3 dimensions leads to the following limit \cite{Askar}
\begin{eqnarray}
 \frac{1}{\Delta t} \left[ \frac{1}{\Delta x^2}+\frac{1}{\Delta y^2}+\frac{1}{\Delta z^2}\right]^{-1} \ge \frac{2\hbar}{m}.
\label{eq:schrodinger-stability1}
\end{eqnarray}
For a cubic grid $\Delta x= \Delta y = \Delta z$ we get
\begin{eqnarray}
\frac{\Delta x^2}{3 \Delta t} \ge \frac{2\hbar}{m}.
\label{eq:schrodinger-stability2}
\end{eqnarray}
As before, defining $\Delta v_{grid}= \Delta x/\sqrt{3}\Delta t$, equation (\ref{eq:schrodinger-stability2}) becomes
\begin{eqnarray}
(\Delta v_{grid})^2 \Delta t \ge  \frac{2\hbar}{m},
\label{eq:schrodinger-stability4}
\end{eqnarray}
or
\begin{eqnarray}
\Delta E_{grid} \Delta t \ge  \hbar.
\label{eq:schrodinger-stability4b}
\end{eqnarray}
The above equation implies that in order for the numerical solution of the Schrodinger equation be stable the grid interval must satisfy the above limit. The above limit is valid for any grid size but for the method to be viable this should particularly be true for the limit $\Delta x$, $\Delta t \rightarrow 0$. As mentioned earlier, convergence is an important condition for a FDM to be viable. Therefore, in the limit $\Delta x$, $\Delta t \rightarrow 0$ we approach the quantum realm and in addition make sure that our difference method is convergent.

\vspace*{2mm}

\textit{Refinement path:} As we approach the limit $\Delta x$, $\Delta t \rightarrow 0$ we should also specify the relationship between $\Delta x$ and $\Delta t$. As we approach the limit $\Delta x$, $\Delta t \rightarrow 0$, the  truncation error approaches zero, and the grid velocity $v_{grid}$ remains constant. Here $ v_{grid}$ is the speed with which numerical information flows on the grid. Assuming numerical causality approaches physical causality in the continuum limit this constant should be the speed of light. As the mesh interval size approaches zero this is the limiting speed with which information can be transmitted on the grid. 
 Therefore, $\Delta x$ and $\Delta t \rightarrow 0$ such that $v_{grid}= c$ and the limits on the length and time intervals are
\begin{eqnarray}
\Delta t \ge \frac{2\hbar}{m c^2},
\label{eq:schrodinger-stability4a}
\end{eqnarray}
or
\begin{eqnarray}
\Delta x \ge 2 \sqrt{3} \lambda_c,
\label{eq:schrodinger-stability4b}
\end{eqnarray}
where $\lambda_c=\hbar/mc$ is the reduced Compton wavelength. The above condition implies that causality requires a limit on the minimum spatial grid interval size to be of the order or greater than the Compton wavelength. Another implication of these limits is that the truncation error in the FDE is $O((\Delta x)^2) \gtrsim O(\lambda_c^2)$. Note that we have assumed a cubic grid to derive this limit and similar limits would be true for $\Delta y$ and $\Delta z$. 

In reference \cite{Soriano:2004}, the stability condition for the Schrodinger equation is derived using the FDTD method. 
 The limit in equation (\ref{eq:schrodinger-fd5}) for that case is $\hbar/m$ instead of $2\hbar/m$. Similarly, other higher order methods \cite{Chan} also lead to stability condition similar to equation (\ref{eq:schrodinger-fd5}). Therefore the stability requirements on the limit on grid interval size for various schemes can be written as 
\begin{eqnarray}
\Delta x \ge a_0 \lambda_c,
\label{eq:schrodinger-stability4a}
\end{eqnarray}
where $a_0$ is a constant of order unity and varies from one explicit scheme to another. Hence, in addition to viewing Compton wavelength as a limitation on measuring the position of a particle it can also be viewed as the minimum grid length that can be resolved in order to ensure causality in the stability analysis of the Schrodinger equation.  Furthermore, assuming a minimum length in this manner renders the theory the same causal structure as the wave equation.

\section{The explicit method and its implications} \label{numerical-discussion}

We have seen in previous sections that the explicit method can shed light on some important ideas in physics.
It is important to note the difference between the explicit and implicit method. The explicit method is fast but less stable and requires small time steps for more accuracy. The implicit method is slower, more stable and allows a larger time-step. In addition to these factors, the explicit method has some important features that makes it different from implicit methods. Following can be the two reasons as to why the explicit method yields these important relationships in contrast to the implicit methods:
\begin{enumerate}
\item The solutions at present time are explicit functions of solutions of past. This implies temporal causality.

\item The solution at a grid location do not instantaneously effect the solutions at the entire grid \cite{Holmes:2006}. This implies a finite speed of information transfer which is called the grid speed. 
\end{enumerate}

We can better understand these points by considering the example of the heat equation which does not limit the speed of heat transmission. However, when using the explicit finite difference method in seeking the numerical solution of a differential equation this might not necessarily hold. The heat equation, for instance, has the instant messaging property but it does not hold when we employ the explicit method to solve it \cite{Holmes:2006}. 
 The speed with which information travels through the grid is $\Delta x/ \Delta t$ and not infinite. This, of course, implies that the explicit method is not the perfect method to solve the heat equation but at the same time it does render the system a causal structure. 
 The implicit method, however, does have the instant messaging property and this is apparently the reason it is unconditionally stable. Therefore, the numerical propagation speed of information for explicit methods is finite whereas that of implicit method is infinite.

 The Schrodinger's equation is similar to the heat equation but with an imaginary diffusion coefficient. It is apparently a parabolic PDE but the solution it permits has properties of the solutions of hyperbolic PDEs.  In other words although it satisfies the mathematical test of a parabolic equation which describes diffusive processes it allows plane wave solutions. Parabolic PDEs allow infinite speed of information transfer whereas this speed is finite for hyperbolic PDEs. In both cases, however, the explicit finite difference method when employed allows only a finite speed of propagation.

\section{Conclusion}\label{conclude}

We discussed how the techniques used in numerical analysis might be more than tools to obtain  solutions of partial differential equations. Finite difference methods and, in particular, explicit methods might help us further understand some fundamental ideas in physics. Explicit methods used in solving finite difference equations are often conditionally stable. The explicit finite difference method for the wave equation can lead to a limit on the speed of waves to be less than the grid speed. As the grid approaches the continuum limit, we assumed that the idea of numerical causality merges with physical causality and
the speed of numerical information transfer becomes equal to the physical speed limit, i.e., the speed of light. 
This leads to a limit on the speed of waves that ensures stability of the explicit method and also ensures physical causality.

 Similarly, the explicit method of the Schrodinger equation yields a limit on the minimum length and time interval.  To ensure convergence of the explicit method, we approach zero interval size along the refinement path $v_{grid}=c$. If we insist on physical causality in the continuum limit this speed should be the limiting speed of information travel, i.e., the speed of light. We thereby showed that the limits on the minimum spatial and temporal grid lengths are $\Delta x \ge 2 \sqrt{3} \hbar/mc$ and $\Delta t \ge  \hbar/mc^2$. This can be understood as follows. In the continuum limit, the grid interval, being a measure of the position of the particle, suffers from similar limitations as those obtained in quantum mechanics. These limits result from the Von Neumann stability analysis. Therefore, the explicit method endows a causal structure to the Schrodinger equation which is similar to that of the wave equation and as a result implies the minimum grid interval size to be of the order of Compton's wavelength to guarantee stability. 

 Finally, we also discussed the importance of the explicit finite difference method in numerical analysis. The Von Neumann stability analysis used to test the stability of explicit methods implies that our conclusions are essentially independent of boundary conditions. The fact that this method preserves causality and has a finite speed of information transfer can be two important reasons it yields important limits on the spatial and temporal grid interval sizes. 
 Therefore, in our analysis we showed that insisting on physical causality and convergence of a finite difference method can lead to important ideas in fundamental physics.

\section{Acknowledgments}
The author would like to thank Fariha Nasir for useful discussions.

\end{document}